\documentclass[journal]{IEEEtran}
\newcommand{\figurewidth}{0.25\textwidth}
\usepackage{cite}
\usepackage{graphicx}
\usepackage{epsfig,psfrag,subfigure}
\usepackage{array}
\usepackage{theorem}
\usepackage{amsmath}
\usepackage{amssymb}

\newtheorem{theorem}{Theorem}

\newtheorem{definition}{Definition}

\newcommand{\mc}{\mathcal}
\newcommand{\mb}{\mathbf}

\newcommand{\FP}{{Frobenius-Perron}  }

\begin{document}
%
\title{A Framework for Investigating the Performance of\\ Chaotic-Map Truly Random Number Generators}
%
%
%


\author{Ahmad~Beirami~\IEEEmembership{Student~Member,~IEEE,}
        and~Hamid~Nejati\vspace{-.02in}


\thanks{A. Beirami is with the School
of Electrical and Computer Engineering, Georgia Institute of Technology, Atlanta,
GA, 30332 USA e-mail: (beirami@ece.gatech.edu).}

\thanks{H. Nejati is with the Department of Electrical Engineering and Computer Science, University of Michigan, Ann Arbor, MI, 48109 USA e-mail: (hnejati@umich.edu).}}


\maketitle
\begin{abstract}
In this paper, we approximate the hidden Markov model of chaotic-map truly random number generators (TRNGs) and describe its fundamental limits based on the approximate entropy-rate of the underlying bit-generation process. We demonstrate that  entropy-rate  plays a key role in the performance and robustness of chaotic-map TRNGs, which must be taken into account in the circuit design optimization. We further derive optimality conditions for post-processing units that extract truly random bits from a raw-RNG.
\end{abstract}
\vspace{-.03in}



\IEEEpeerreviewmaketitle

\vspace{-.05in}
\section{Introduction}
\label{sec:intro}

\IEEEPARstart{C}{haos} is the long-term non-predictive behavior observed in certain nonlinear dynamical systems due to the sensitivity of the output trajectory to the initial conditions. Although chaotic systems are deterministic and can be described using simple differential equations, their output trajectory can only be determined given the {\em exact} initial state of the system. As in practice the initial state of the system is only known with {\em uncertainty} due to the random environmental noise, the perturbations grow exponentially leading to unpredictability~\cite{applied_chaos}.

One natural application for the unpredictability in chaotic systems is in random number generators (RNGs).
A RNG requires a \emph{never-ending} source of entropy (randomness), which can be supplied by the inherent natural noise in the analog circuitry~\cite{holman97,bucci03,petrie00,chaotictelecomm2000}. One of the most prominent solutions for random number generation is based on the chaotic maps~(cf.~\cite{wang05a,addabbo06a,ALOG12,callegari05}).
A chaotic-map RNG operates based on the amplification of the inherent noise in a chaotic map function by feeding the output back into the map in each time step~\cite{callegari05}. Then, the map output is transformed into a binary random variable using a bit generation function. Chaotic-map RNGs are fast and easily integrable, and hence, interesting in practice. 

An \emph{ideal} chaotic-map RNG should be capable of generating a \emph{truly random} bit sequence, which consists of \emph{independent} and \emph{equiprobable} output symbols (bits).
Truly random number generators (TRNGs) are required in public key cryptography as well as digital signature schemes as essential tools for data protection~\cite{crypto_handbook}.
On the other hand, due to the process variations, there is an inevitable correlation in the generated output bits in a practical RNG (hereafter called raw-RNG). An important question is how robust the map function is to implementation non-idealities and how could the robustness be measured.
The degraded performance due to the correlations necessitates the utilization of a post-processing unit for the recovery of the randomness in the bit sequence generated by a given raw-RNG so as to obtain a TRNG (cf.~\cite{von-neumann51, addabbo06b}).


In~\cite{callegari05}, the authors presented a mathematical analysis of  piece-wise affine map-based raw-RNGs using a Markov model. In this paper, we describe the shortcomings of the analysis in~\cite{callegari05} and present a more comprehensive analysis.
Our contributions are as follows:
\begin{itemize}
\item We derive the approximate entropy-rate of the hidden Markov process underlying chaotic-map raw-RNGs to determine their fundamental limits.

\item We demonstrate the effectiveness of the entropy-rate in investigating the robustness of chaotic-map raw-RNGs.

\item We derive necessary (though not sufficient) conditions that an ideal post processing unit should fulfill to extract truly random bits from a raw-RNG.
\end{itemize}

\begin{figure}
\centering
\vspace{0.05in}
\includegraphics[height=0.4\linewidth,angle=-90]{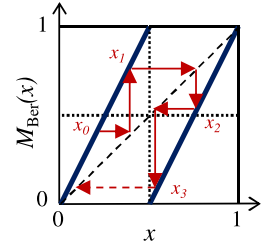}
\vspace{-0.07in}
\caption{The Bernoulli shift map.}
\vspace{-0.15in}
\label{fig:maps}
\end{figure}

\begin{figure*}
\vspace{-0.05in}
\begin{center}
  \subfigure[~]{
      \psfrag{xaxis}{\small{$x$}}
    \psfrag{yaxis}{\small{\hspace{-.2in}$M_\text{ex}(x)$}}
    \psfrag{Y: 0.4}{\tiny{$~$}}
\psfrag{X: 0.1065}{\tiny{$u_1$}}
\psfrag{X: 0.5463}{\tiny{$u_2$}}
  \includegraphics[width=\figurewidth]{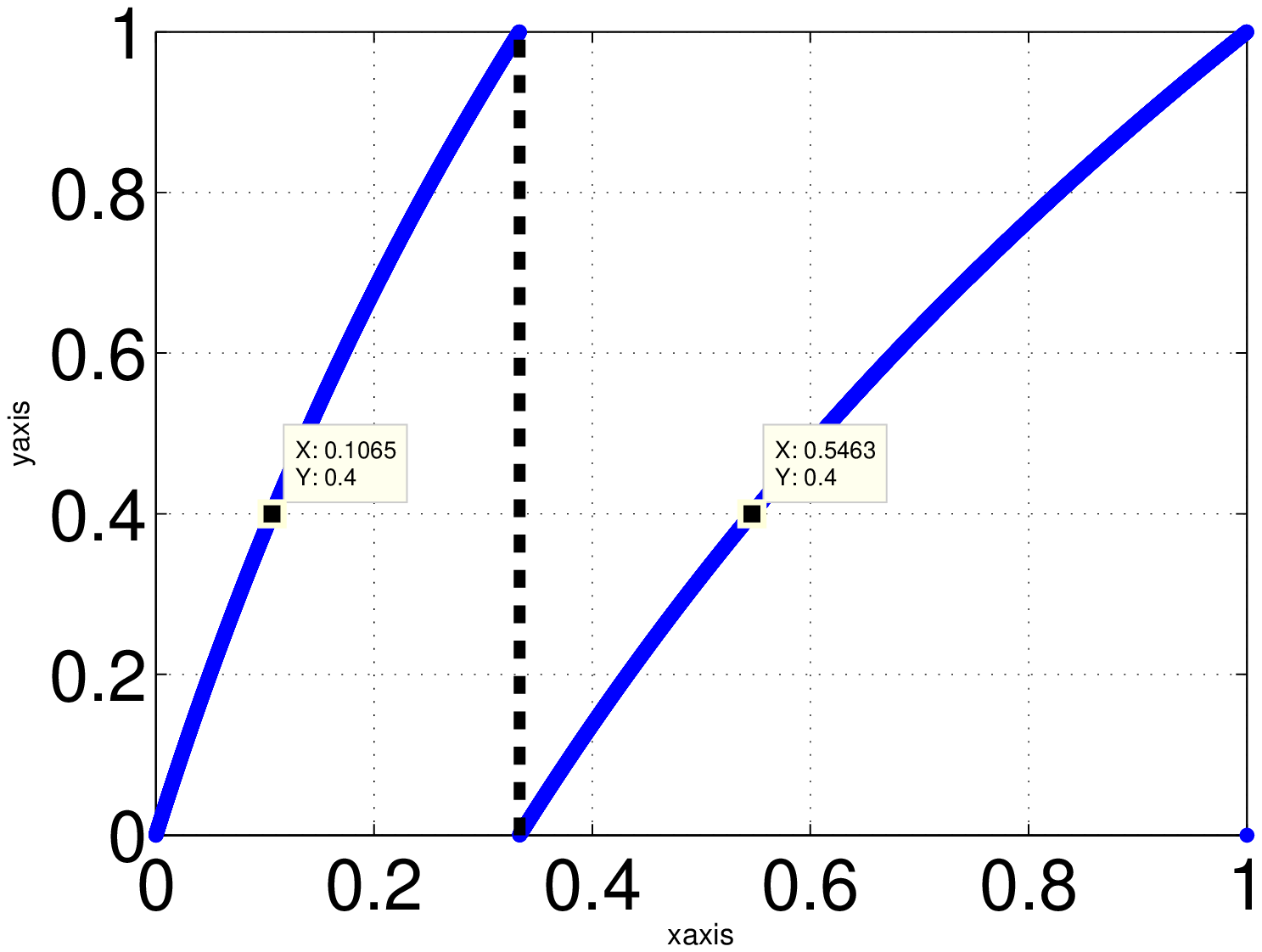}
    \label{fig:sample_TCAS2_map}
  }
  \hspace{0.03\textwidth}
  \subfigure[~]{
      \psfrag{xaxis}{\small{$x$}}
    \psfrag{yaxis}{\hspace{-.15in}\small{$f_\infty(x)$}}
    \psfrag{Y: 0.3763}{\tiny{$~$}}
\psfrag{Y: 0.8054}{\tiny{$~$}}
\psfrag{X: 0.1065}{\tiny{$u_1$}}
\psfrag{X: 0.5465}{\tiny{$u_2$}}
  \includegraphics[width=\figurewidth]{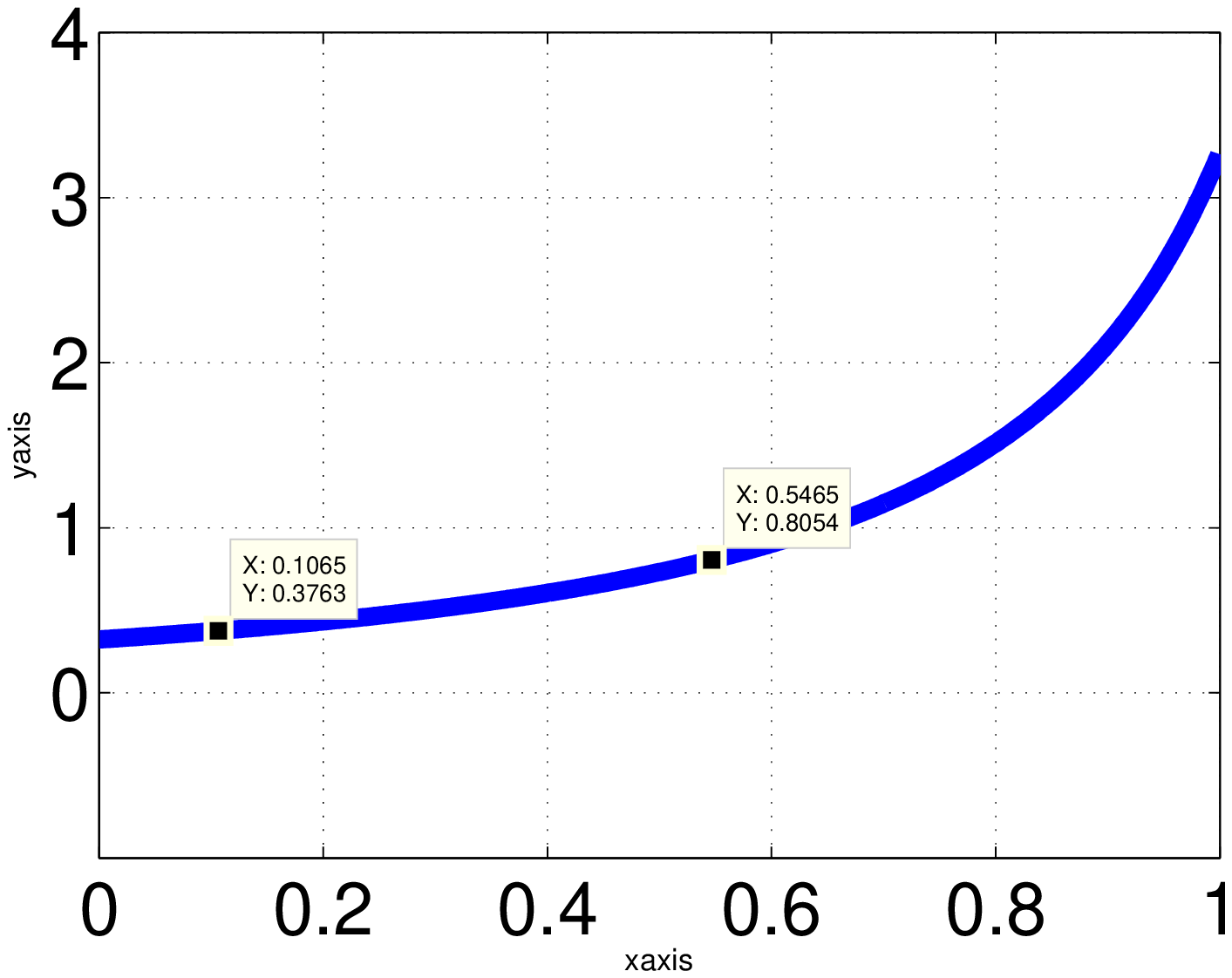}
    \label{fig:sample_TCAS2_density}
  }
  \hspace{0.03\textwidth}
  \subfigure[~]{
     \psfrag{xlabel}{\small{$n$}}
    \psfrag{ylabel}{\hspace{-0.3in}\small{$H(Z_n|Z^{n-1})$}}
    \psfrag{Y: 0.4}{\tiny{$y$: $0.4$}}
\psfrag{X: 0.1065}{\tiny{$x_1$: $0.107$}}
\psfrag{X: 0.5463}{\tiny{$x_2$: $0.546$}}
  \includegraphics[width=\figurewidth]{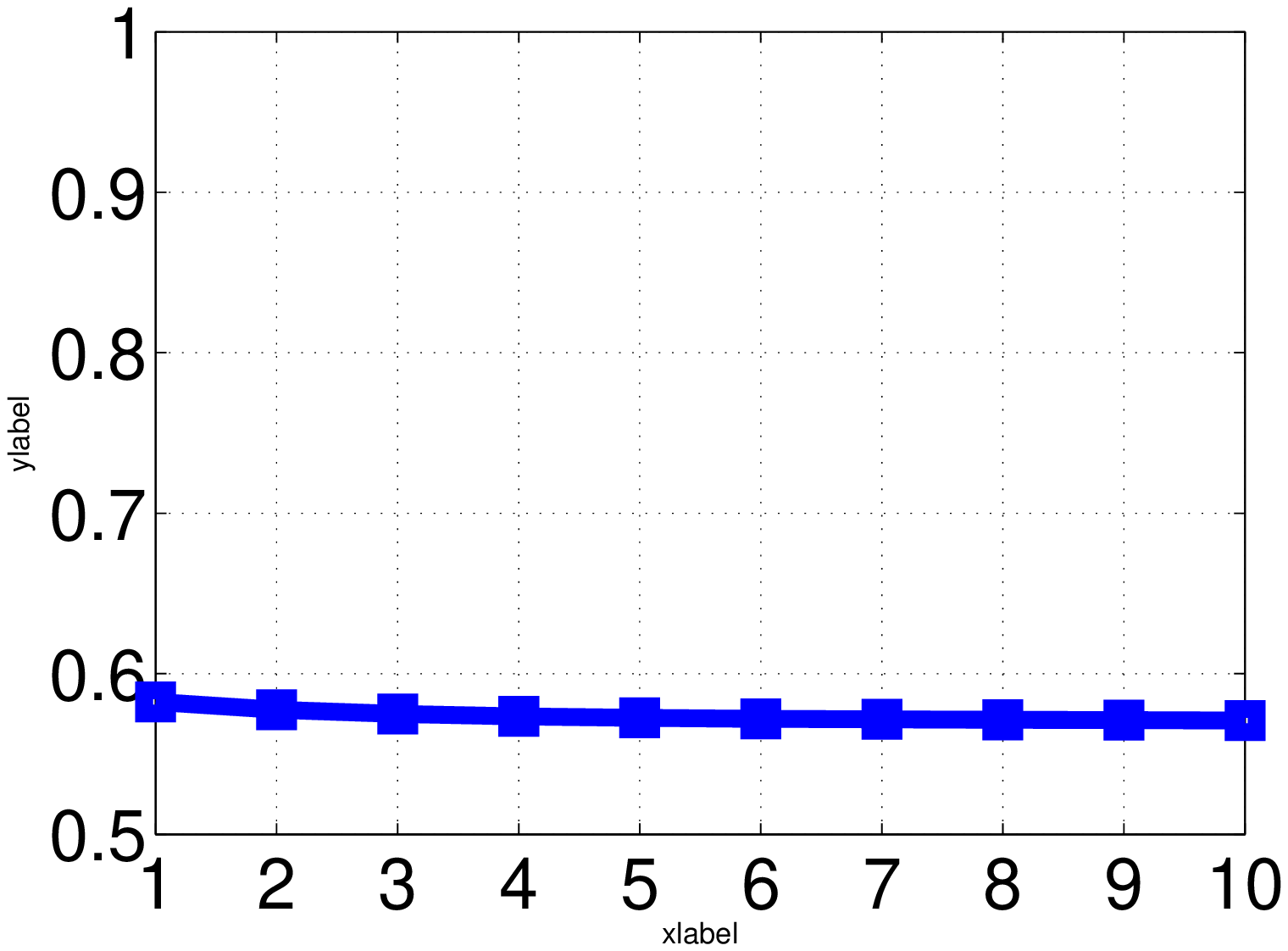}
    \label{fig:sample_TCAS2_entropy}
  }
\end{center}
\vspace{-.12in}
\caption{(a) The example map. (b)~Steady-state probability distribution. (c)~Conditional entropy.}
\label{fig:sample_TCAS2_ALL}
\vspace{-.15in}
\end{figure*}

\vspace{-.1in}
\section{Background Review}
\label{sec:background}
A discrete-time (chaotic) map is defined using a transformation function $M(x): (0,1) \rightarrow (0,1)$. The output of the map function from the previous time-step is fed back as the input of the current time-step. That is
$
x_{n} = M(x_{n-1}) = M^{n-1}(x_0),
$
where $M^{i}(x_0) \triangleq  M(M^{i-1}(x_0))$, $n$ is the time step, $x_0$ is the initial state of the system, and $x_n$ is the state of the system at time $n$, and clearly, the sequence $\{x_n\}$ forms a Markov process~\cite{applied_chaos}.
Let $f_n(\cdot)$ denote the probability density function (PDF) of the map output at time $n$. Due to the Markov property, the probability distribution at time $n$  is related to that of time $(n-1)$ through the linear \FP operator $P$ as given by $f_{n}(x) = P f_{n-1} (x)$.
If the initial state of this deterministic system (i.e., $x_0$) does not follow a PDF of bounded variation (e.g., is {\em exactly} known), the output behavior can be predicted~\cite{chaosfractal94}.
However, $f_0(x)$ is of bounded variation as a result of random perturbations, and hence the corresponding Markov chain is ergodic. Hence, $f_n(x)$ asymptotically approaches a fixed steady-state distribution denoted by $f_\infty(x)$ when $n \rightarrow \infty$, which satisfies $f_\infty(x) = Pf_\infty(x)$.
Let the Lyapunov exponent associated with a discrete-time map be defined as $\lambda = \int_0^1 \ln  |M'(x)| f_\infty(x) dx$, where $M'(x)$ is the derivative of $M(x)$. We assume that $M(x)$ is continuous and differentiable except in finitely many points. The discrete-time system defined above exhibits chaotic behavior if $\lambda>0$. 


Next, we briefly review the Bernoulli shift map (shown in Fig.~\ref{fig:maps}), which is a piecewise-affine function given by
\begin{equation}
M_\text{Ber}(x) = \left\{  \begin{array}{ll}
2x & 0<x<\frac{1}{2}  \\
2x-1 &  \frac{1}{2}<x<1
\end{array}
\right..
\end{equation}
The Lyapunov exponent of the map is equal to $\ln(2)$.
The \FP operator $P$ is implicitly given by
$f_{n}(x)\hspace{-.02in} =\hspace{-.02in} Pf_{n-1}(x)\hspace{-.01in} =\hspace{-.01in} \frac{1}{2}\hspace{-.02in}\left\{ \hspace{-.01in} f_{n-1}\left(\frac{x}{2}\right) \hspace{-.02in}+ \hspace{-.02in}f_{n-1}\left( \frac{1}{2}+\frac{x}{2} \right)\hspace{-.03in} \right\}\hspace{-.03in}.
$
It is straightforward to show that $f_\infty(x)=1$ is a solution to $f_\infty = P f_\infty$, and hence, the steady-state probability distribution for the Bernoulli map is uniform over $(0,1)$.

A random number generator that operates based on a chaotic map can be defined using the pair $(M,\mc{B})$ where $M: (0,1) \to (0,1)$ is the map function and $\mc{B}: (0,1) \to \{0,1\}$ is the bit-generation function.
Let the output binary sequence be denoted by $\{z_n\}$, where
$
z_{n} \triangleq \mc{B}(x_{n-1}).
$
We assume that the initial state of the system follows the steady-state distribution, i.e., $X_0$ follows
$f_\infty$.\footnote{In this paper we use the upper case letter to denote a random variable and the lower case letter to denote a realization of the random variable.}
Consequently, the output sequence $\{x_n\}$ of the raw-RNG forms a stationary Markov process.
Further, as the user does not have access to the state sequence $\{x_n\}$ of the chaotic map, the sequence $\{z_n\}$ by definition forms a hidden Markov process~\cite{HMP_IT,Myers92}.  A hidden Markov process is a random process whose outcomes are functions of a Markov process.

For the case of the Bernoulli shift map, a raw-RNG will be formed using the pair ($M_\text{Ber}, \mc{B}_\text{Ber})$, where $\mc{B}_\text{Ber}$ is defined as
\begin{equation}
z_{n} =\mc{B}_\text{Ber}(x_{n-1}) \triangleq  \lfloor x_{n-1} + \frac{1}{2} \rfloor,
\label{eq:bit_generation}
\end{equation}
where $\lfloor x \rfloor$ denotes the greatest integer less than or equal to $x$.
In other words, the output is $z_n= 0$  if $x_{n-1}<1/2$ and $z_n = 1$ if $x_{n-1}\geq 1/2$.
It is straightforward to verify that the bits generated from the Bernoulli shift map raw-RNG are independent and equally $0$ or $1$  and hence {\em truly random}, i.e., we can show that $\mb{P}[z_{n}|z^{n-1}] = \mb{P}[z_{n}] = \frac{1}{2}$.\footnote{For any $i\leq j$, we define $z_i^j \triangleq (z_i,z_{i+1}\ldots,z_j)$ as the sequence of all binary outputs from time $i$ to time $j$. We further denote $z^j \triangleq z_1^j$.}

In~\cite{callegari05}, the authors presented a mathematical analysis of piece-wise affine map-based raw-RNGs using a Markov model and applied it to a variant of the Bernoulli shift map. Since the hidden Markov bit-generation process does not generally yield to a Markov process, the model in~\cite{callegari05} is not capable of explaining the bit-generation process and investigating the robustness of the map to implementation variations for a general chaotic map.

\vspace{-.05in}
\section{Statistical Properties of Bit-Generation}
\label{sec:binary_sequence}
In this section, we theoretically formulate the bit-generation process from a general discrete-time chaotic map.
Let $N(x)$ denote the number of solutions to the equation $M(u) = x$. Further, let $u_1, u_2, ..., u_{N(x)}$ be the $N(x)$ solutions. For example, it is seen that in the Bernoulli map we have $N(x) = 2$ for all $x \in (0,1)$ except at the break-points.
The \FP operator for the chaotic map is then expressed as
$
f_{n}(x) =  P f_{n-1}(x) = \sum_{i=1}^{N(x)}\frac{1}{|M'(u_i)|}f_{n-1}(u_{i}).
\label{eq:FP_sample_map}
$

In order to clarify the discussions, we consider the fairly general example map function shown in Fig.~\ref{fig:sample_TCAS2_map}. Note that, however, the framework is generic and is applicable to any chaotic map. The example map function $M_\text{ex}$ is given by
$
M_\text{ex}(x)\hspace{-.01in} =\hspace{-.01in} \log_2 (1+3x) \hspace{-.02in}- \hspace{-.02in}\lfloor\log_2 (1+3x) \rfloor\hspace{-.02in}.
$
Further, let the bit generation function for the example map be chosen to be
$z_n = \mc{B}_\text{ex}(x_{n-1}) = \lfloor x_{n-1}+\frac{2}{3} \rfloor.
$
Thus, $z_n = 1$ ($z_n=0$) whenever the input is $x_n \geq \frac{1}{3}$ ($x_n < \frac{1}{3}$), as shown by the vertical dashed line in Fig.~\ref{fig:sample_TCAS2_map}.
It is easily seen that there are exactly two solutions to  $M(u) = x$ for all $x$  in the example map (except at the break-points). Note that this need not be the case in a general map. For example, at $x=0.4$, the steady-state distribution is dependent on the steady-state distribution at points $u_1$ and $u_2$ as shown in the figure.
The solution to the steady-state probability distribution for the example map is illustrated in Fig.~\ref{fig:sample_TCAS2_density}.



Let the set $S_n(z^n) \subset (0,1)$ be defined as follows.
\begin{equation}
S_n(z^n) \triangleq \left\{x_0\left|~\begin{array}{l}
\mc{B}(x_0) = z_1\\
\mc{B}(x_1)=z_2\\
~~\vdots \\
\mc{B}(x_{n-1}) = z_n
\end{array}
\right.
\right\},
\end{equation}
where $x_j$ for all $0\leq j\leq n-1$ denotes the state of the system at time $j$.
In other words, $S_n(z^n)$ is the set of all initial points $x_0$ for which the first $n$ generated bits are the bit sequence $z^n$.
Thus, given $S_n(z^n)$ , we determine the probability of the event that the sequence $z^n$ is generated, as given by
\begin{equation}
\mb{P}[z^n] = \int_{x \in S_n(z^n)} f_\infty(x) dx.
\end{equation}

As a special case, $S_1(z_1) = \{x_0|\mc{B}(x_0) = z_1\}$ is the set of all points in the map that generate the output bit $z_1$ (either $0$ or $1$). This is indeed the definition of the bit-generation function. Accordingly, we can calculate the probability of the event that $z_1$ is generated by $\mb{P}[z_1] = \int_{x \in S_1(z_1)} f_\infty(x) dx$.
This calculation leads to obtaining the widely used measure for the randomness of a binary sequence, i.e., the bias $b$, as
$
b \triangleq  \left| \mb{P}[0]-\frac{1}{2}\right| = \left|\mb{P}[1]-\frac{1}{2}\right|.
$
The bias is a measure of the closeness of the average number of $1$'s in the sequence to the desired value of $\frac{1}{2}$. In the case where the generated bits in the output sequence are known to be independent, {\em bias} is the only determining factor for the randomness of the sequence, whereas a zero bias ($b=0$) indicates that the binary sequence is in fact truly random.
On the other hand, as we shall see shortly, the hidden Markov process defined here to investigate process variations oftentimes entails correlation between the bits in the sequence, i.e., the knowledge of $z_i$ will provide partial information about whether $z_j$ is $0$ or $1$ for all $i<j$.

To clarify, we again consider the example map in Fig.~\ref{fig:sample_TCAS2_map}. Here, by definition of $\mc{B}_\text{ex}$, we have $S_1(0) = (0,\frac{1}{3})$, and $S_1(1) = (\frac{1}{3},1)$.
Accordingly, we can determine $\mb{P}[0] = 0.14$, $\mb{P}[1] = 0.86$, and the bias to be $b = 0.36$.
This indicates that the generated bit sequence cannot be truly random.

As described above, we need to determine $S_n(z^n)$ in order to calculate $\mb{P}[z^n]$. We have
\begin{equation}
x_0 \in S_{n}(z^{n}) \iff
\left\{
\begin{array}{l}
x_0 \in S_1(z_1)\\
x_1 \in S_{n-1}(z_2^n)
\end{array},
\right.
\label{eq:partition}
\end{equation}
where $x_1 = M(x_0)$.
Thus, $S_{n}$ is obtained recursively using $S_{n-1}$ and $S_1$ assuming that $S_{n-1}(z^{n-1})$ is known for all possible $2^{n-1}$ binary sequences of length $n-1$.

\section{Fundamental Performance Limits}
\label{sec:limits}
In this section, we derive the entropy-rate of the bit-generation process and analyze the fundamental performance limits of TRNGs.
As a chaotic-map raw-RNG is governed by a hidden Markov process,  no closed form solution exists for the entropy-rate in general~\cite{HMP_IT}. Thus, an approximate solution is needed. Let the entropy of the sequence $z^n$ be defined as~\cite{Cover_book}
\begin{equation}
H(Z^n) = \sum_{z^n \in \{0,1\}^{n}} \mb{P}[z^n] \log_2\left( \frac{1}{\mb{P}[z^n]}\right).
\end{equation}
Let $H(Z_n|Z^{n-1})$ denote the conditional entropy of the random symbol $Z_n$ given the latest $n-1$ symbols (i.e., $Z^{n-1}$). That is
$
H(Z_n|Z^{n-1}) = H(Z^n) - H(Z^{n-1})$~\cite{Cover_book}.
It is straightforward to show that $H(Z_n|Z^{n-1})$ is a monotonically decreasing function of $n$, and hence, $H(Z_n|Z^{n-1})$ converges to a fixed value $H$ called the {\em entropy-rate} of the generated binary sequence as given by~\cite{Cover_book}
\begin{equation}
H \triangleq \lim_{n \rightarrow \infty}{H(Z_n|Z^{n-1})} = \lim_{n \to \infty}\frac{1}{n} H(Z^n).
\label{eq:entropy}
\end{equation}
Note that since $\{Z_n\}$ comes from a hidden Markov process in the case of chaotic-map raw-RNGs, the rate of convergence of $H(Z_n|Z^{n-1})$ to the entropy-rate $H$ is exponential in $n$, where the exponent is proportional to the Lyapunov exponent $\lambda$ of the underlying chaotic map~\cite{HMP_IT}.

Let $\mc{P}_{n}: \{0,1\}^{n} \to \{0,1\}^{k(n)}$ denote a {\em deterministic} post-processing function such that it inputs a random binary sequence $z^n$ and outputs another random sequence denoted by $t^k= \mc{P}_{n}(z^n)$.   Further, let $R(\mc{P}) \triangleq \lim_{n \to \infty} \frac{k(n)}{n}$ denote the post-processing rate. Please note that for any $n$, the output rate of the post-processing unit can take $(n+1)$ discrete values of $\left\{0,\frac{1}{n},\ldots,1\right\}$ based on the value of $k(n)$.
\begin{definition}
A post-processing function $\mc{P}_{n}$ is {asymptotically truly random} if $\lim_{n\to\infty}\frac{1}{k}H(T^k) = 1.$
\label{def:truly-random}
\end{definition}
Intuitively, a truly random post-processing unit is one that asymptotically for large $n$ turns the input sequence to a nearly truly random (independent and equiprobable) bit sequence.

\begin{figure*}
\vspace{-0.05in}
\begin{center}
  \subfigure[~]{
      \psfrag{xlabel}{\small{$x$}}
    \psfrag{ylabel}{\hspace{-0.15in}\small{$M(x)$}}
  \includegraphics[width=\figurewidth]{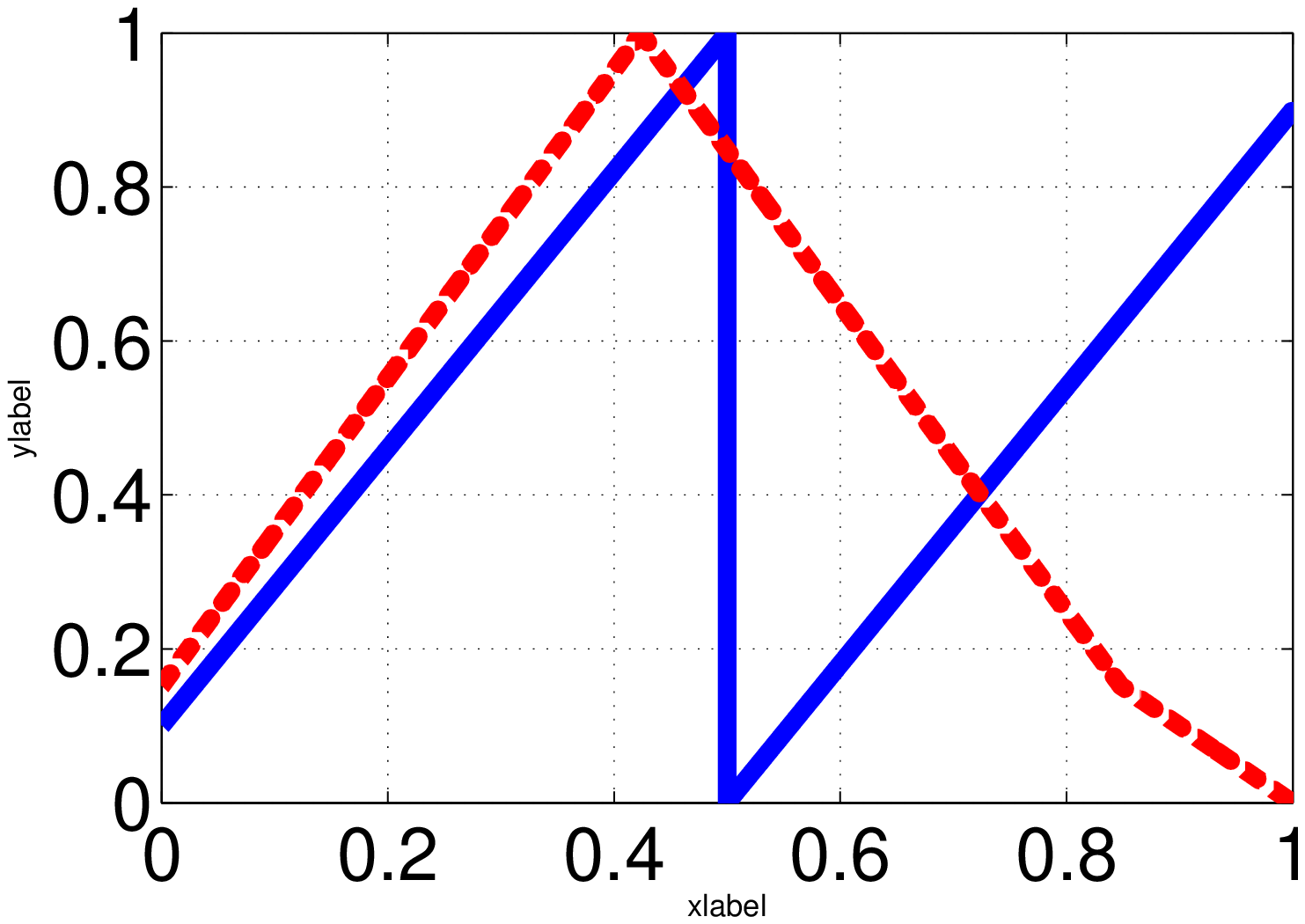}
    \label{fig:ex_TCAS2_map}
  }
  \hspace{0.03\textwidth}
  \subfigure[~]{
      \psfrag{xlabel}{\small{$x$}}
    \psfrag{ylabel}{\hspace{-0.15in}\small{$f_\infty(x)$}}
  \includegraphics[width=\figurewidth]{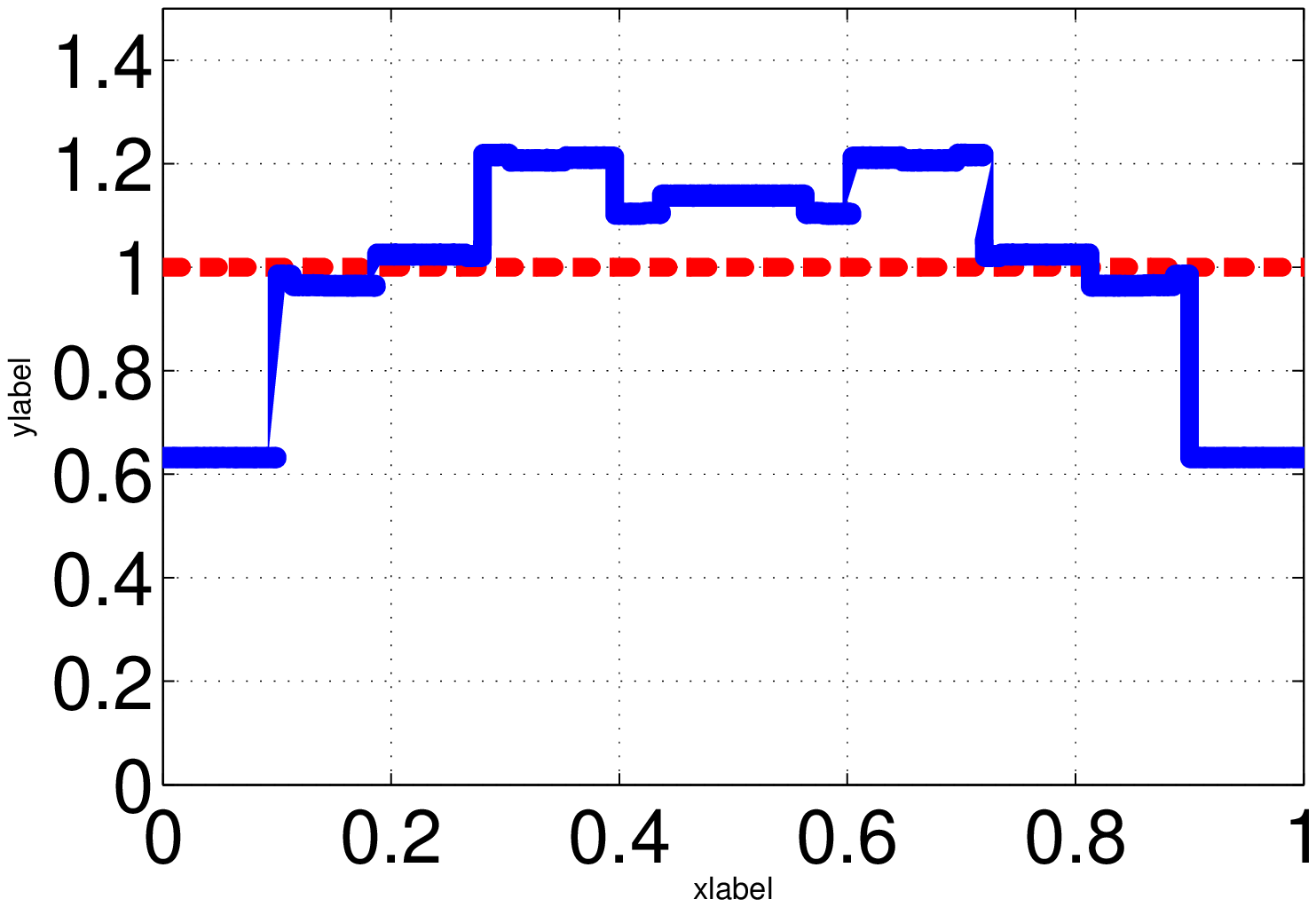}
    \label{fig:ex_TCAS2_density}
  }
  \hspace{0.03\textwidth}
  \subfigure[~]{
     \psfrag{xlabel}{\small{$n$}}
    \psfrag{ylabel}{\hspace{-0.3in}\small{$H(Z_n|Z^{n-1})$}}
  \includegraphics[width=\figurewidth]{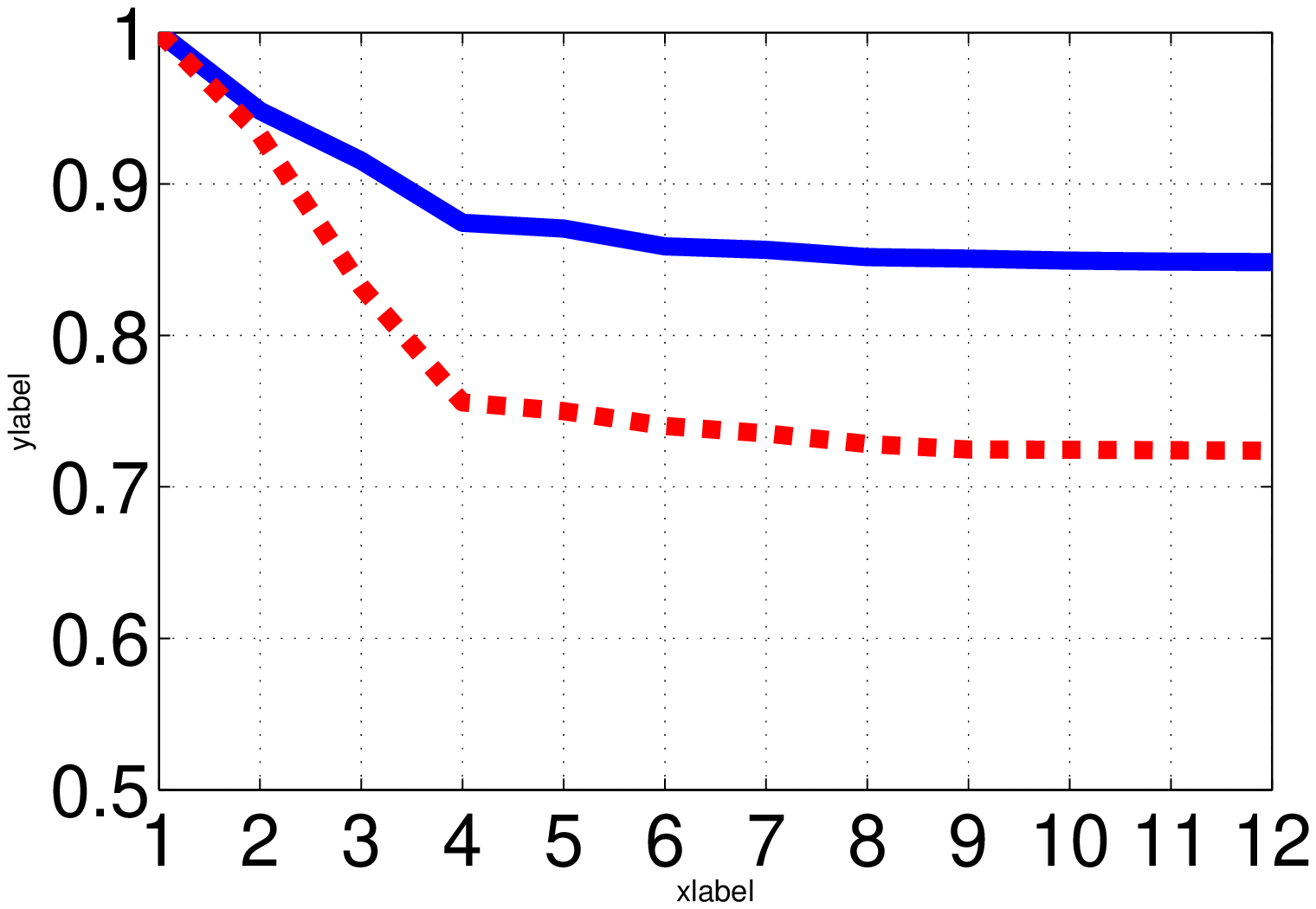}
    \label{fig:ex_TCAS2_entropy}
  }
\end{center}
\vspace{-.15in}
\caption{The solid blue curves correspond to the decreased slope Bernoulli shift map. The dashed red curves correspond to the tailed tent map. (a) Decreased slope Bernoulli shift map function: $M_\text{Dec-Bernoulli}(x)$ (solid blue curve) and tailed tent map function: $M_\text{Tailed-Tent}(x)$ (dashed red curve). (b)~Steady-state probability distribution.~(c) Conditional entropy.}
\label{fig:ex_TCAS2_ALL}
\vspace{-.13in}
\end{figure*}

Next, we will present our main results that will be used to determine the necessary but not sufficient condition for ideal performance evaluation of the chaotic-map TRNGs.
\vspace{-.07in}\begin{theorem}
If the post-processing function $\mc{P}_{n}$ is asymptotically truly random, then we must have $R(\mc{P})\leq H$.
\label{thm:UB}
\end{theorem}
\vspace{-.05in}
\begin{IEEEproof}
We present a proof by contradiction. Assume that $R(\mc{P})>H$. According to the data processing theorem~\cite{Cover_book}, we have $H(T^k) \leq H(Z^n)$. Hence,
\begin{equation}
\lim_{n\to\infty} \frac{1}{k}H(T^k) \leq \lim_{n\to\infty}\frac{1}{k}H(Z^n) = \frac{H}{R(\mc{P})} <1,
\end{equation}
where the equality is due to the fact that, by definition, $\lim_{n \to\infty}\frac{k}{n} = R(\mc{P})$ and $\lim_{n\to\infty}\frac{1}{n}H(Z^n) = H$.
\end{IEEEproof}

Theorem~\ref{thm:UB} states that unless we do not {\em wisely} decrease the rate by a factor $H$, we cannot turn the output bits of the raw-RNG to a truly random sequence.
Again, considering the example map in Fig.~\ref{fig:sample_TCAS2_map}, the entropy-rate can be shown to be $\sim0.57$ (Fig.~\ref{fig:sample_TCAS2_entropy}), i.e., the rate of a truly random post-processing unit is smaller than $0.57$.
\vspace{-.07in}\begin{theorem}
For any  $R\leq H$, there exists an asymptotically truly random  post-processing function $\mc{P}_{n}$ with $R(\mc{P}) = R$.
\label{thm:LB}
\end{theorem}
\vspace{-.05in}
\begin{IEEEproof}
It suffices to demonstrate that  a truly random post-processing with asymptotic rate $H$ exists. Then, any rate $R\leq H$  is also achieved by simply discarding $\frac{R}{H}$ of the bits in the sequences of length $nH$.
To show that rate $H$ is achievable, we build upon the idea of universal source coding and the asymptotic equipartition property~\cite{Cover_book}.
For any $\epsilon >0$, let $A_\epsilon^{(n)}$ denote the {\em typical set} with respect to $\mb{P}[Z^n]$, i.e.,~\cite{Cover_book}
\begin{equation}
A_\epsilon^{(n)} = \left\{z^n \in \{0,1\}^n\left|2^{-n(H+\epsilon)}\leq \mb{P}[Z^n] \leq 2^{-n(H-\epsilon)}\right.\right\}.\nonumber
\end{equation}
 Then, using the Shannon-McMillan-Breiman theorem~\cite{Shannon-McMillan}, it is shown that $\mb{P}[Z^n \in A_\epsilon^{(n)}]\geq 1-\epsilon$. For sufficiently large $n$, we also have $ (1-\epsilon)2^{n (H - \epsilon)} \leq |A_\epsilon^{(n)}| \leq 2^{n(H + \epsilon)}$~\cite{Cover_book}. In other words, any sequence of length $n$, with probability at least $(1-\epsilon)$, is contained in the typical set and the size of the typical set is no larger than $2^{n(H + \epsilon)}$. Let $k= \log  |A_\epsilon^{(n)}|$.
 Label all sequences $z^n \in A_\epsilon^{(n)}$ with binary sequences $t^{k}(z^n)$ of length $k$ bits. Denote $\mb{0}^k$ as the all-zero sequence of length $k$. Then, let the post-processing function $\mc{P}_{n}(\cdot)$ be defined by
 \begin{equation}
 \mc{P}_{n}(z^n) =
 \left\{
 \begin{array}{ll}
 t^{k}(z^n)&\textit{if $z^n\in A_\epsilon^{(n)}$}
 \\
 \mb{0}^k&\textit{otherwise}
 \end{array}
 \right.
 ,
 \label{eq:pp-optimal}
 \end{equation}
 Define $\mathbb{I}_{z^n}$ as the indicator of the typical set such that $\mathbb{I}_{z^n} =1 $ if $z^n\in A_\epsilon^{(n)}$ and  $\mathbb{I}_{z^n} =0 $ otherwise. By the properties of entropy~\cite{Cover_book}, we have $H(T^k)\geq  H(T^k|\mathbb{I}_{z^n})$. Further,
 \begin{equation}
H(T^k|\mathbb{I}_{z^n}\hspace{-.02in})\hspace{-.02in}=\hspace{-.02in}H(T^k|\mathbb{I}_{z^n}\hspace{-.05in}=\hspace{-.05in}1)\mb{P}[\mathbb{I}_{z^n} \hspace{-.05in}=\hspace{-.05in}1] + H(T^k|\mathbb{I}_{z^n} \hspace{-.05in}=\hspace{-.05in} 0) \mb{P}[\mathbb{I}_{z^n}\hspace{-.05in}=\hspace{-.05in}0]\nonumber,
 \end{equation}
 which is equal to $k\times(1-\epsilon) + 0\times\epsilon = k(1-\epsilon) $, and hence $\frac{1}{k}H(T^k) \geq (1-\epsilon)$.
The post-processing rate is asymptotically $R(\mc{P})= \lim_{n \to \infty }\frac{\log |A_\epsilon^{(n)}|}{n} \geq H  - \epsilon$.
 Since, the above results hold for all $\epsilon>0$, we can choose to let $\epsilon \to 0$, and  we obtain $\frac{1}{k}H(T^k) \to 1$ and $R(\mc{P}) \to H$. Therefore, the post-processing function $\mc{P}_n(\cdot)$ in~(\ref{eq:pp-optimal}) is asymptotically truly random (according to Definition~\ref{def:truly-random}) with rate $H$ as desired.
\end{IEEEproof}


\vspace{-.1in}
\begin{definition}
A post-processing function $\mc{P}_{n}$ is asymptotically optimal if we have $\lim_{n\to\infty} \frac{1}{k}H(T^k) = 1$ and $R(\mc{P}) = H$.
\vspace{-.03in}
\end{definition}
Please note that, among the asymptotically {\em truly random} post-processing units, it is desirable that the rate of post-processing is highest possible, i.e., {\em optimal-rate}. According to Theorem~\ref{thm:LB}, an asymptotically  optimal post-processing unit exists, however, it remains an open problem to provide one.

Although the entropy calculation entails exponential complexity in the sequence length $n$, the convergence to the entropy-rate is also exponential with $n$, where the exponent is proportional to the Lyapunov exponent of the map. Thus far, we have used this method for calculating the entropy-rate of several practical maps and their non-ideal versions (whose Lyapunov exponents are similar to the original map). In the experiments, we have never observed the need to consider lengths beyond $n=10$ for which the entropy calculation takes less than a second using MATLAB on a typical PC.

Note that it is also possible to approximately {\em estimate} the entropy by observing a sufficiently long sequence from a raw-RNG regardless of whether the map function is known or not (as it is done in many randomness test suites)~\cite{Shannon-McMillan}. However, when the map is known, our method of calculation offers two advantages. First, it is significantly faster since estimation requires processing of about one million generated raw bits to be able to give similar predictions as our calculation for $n=10$.
Second, the result of the approximate entropy test is probabilistic, i.e., if a sequence passes the test there is no guarantee that the generator is truly random whereas a sequence with unit entropy-rate is guaranteed truly random. 

\vspace{-.1in}
\section{Significance of the Results}
\label{sec:results}

\vspace{-.05in}
\subsection{Robustness/Randomness Trade-off}
\label{subsec:modified_maps}
To proceed, we utilize the developed methodology for the performance evaluation of two practical robust chaotic maps based on the Bernoulli  map and the tent map. Tent map is a chaotic map with very similar characteristics to those of the Bernoulli map presented in Sec.~\ref{sec:background}. The circuit implementations of both the Bernoulli map and the tent map suffer from the saturation problem, where the output can be saturated in the corner points due to implementation variations, and the output sequence will be all-zeros or all-ones~\cite{chaotictelecomm2000}.

In~\cite{wang05a}, the authors presented a decreased slope Bernoulli map that makes the chaotic map robust by decreasing the slope. The map function for the slope of $1.5$ is shown by the solid blue curve in Fig.~\ref{fig:ex_TCAS2_map}.  As can be seen in Figure~\ref{fig:ex_TCAS2_density} (solid blue curve), the steady-state probability distribution is not uniform. If the bit-generation function is $\mc{B}_\text{Dec-Bernoulli}(x) = \lfloor x+\frac{1}{2} \rfloor$, because of the symmetry of the map and the steady-state distribution the output bit sequence will be unbiased.
Figure~\ref{fig:ex_TCAS2_entropy} (solid blue curve) demonstrates the entropy-rate of the binary sequence output of the decreased slope Bernoulli shift map. As can be seen, despite the unbiased output, the entropy rate is $H \approx 0.84$.


Next, we study the tailed tent map that also resolves the saturation problem~\cite{callegari97}, as shown by the dashed red curve in Fig.~\ref{fig:ex_TCAS2_map}.
The steady-state distribution has been proven to be uniform for all tail parameters, as also confirmed in Fig.~\ref{fig:ex_TCAS2_density}. For the sake of comparison, we chose the tail parameter for which the Lyapunov exponent of the map is equal to the decreased slope Bernoulli map.
The tailed tent map along with the bit-generation function $\mc{B}_\text{Tailed-Tent}(x) = \lfloor x+\frac{1}{2} \rfloor$ can result in a random number generator.
First thing to note is that the output bit stream is unbiased due to the uniform steady-state distribution. However, the conditional entropy converges to a value of $H \approx 0.72$ for the chosen tail parameter (Fig.~\ref{fig:ex_TCAS2_entropy}).
Hence, despite the uniform steady-state distribution and unbiased bit sequence, the performance is even worse than the decreased slope Bernoulli map. 


\vspace{-.05in}
\subsection{Investigating the Robustness to Process Variations}
\label{subsec:zigzag}
\begin{figure}
\centering
\vspace{0.03in}
 \subfigure[~]{
\includegraphics[height=0.31\linewidth]{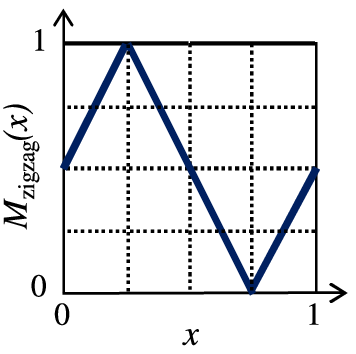}}
\hspace{0.02in}
\subfigure[~]{
\psfrag{ylabel}{\hspace{-.3in}\scriptsize{Perc. of Occur. (\%)}}
\psfrag{xlabel}{\scriptsize{$H$}}
\includegraphics[height= 0.31\linewidth]{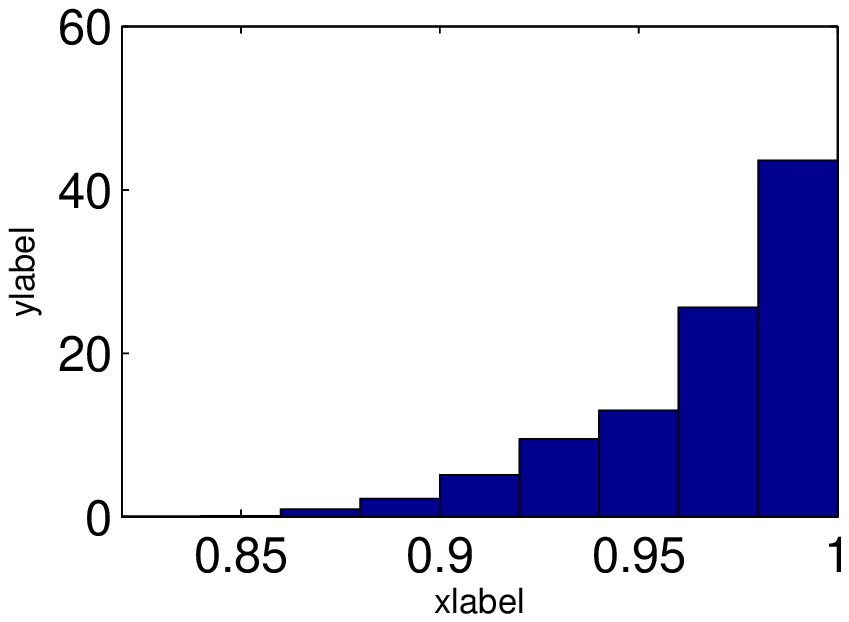}
}
\vspace{-0.05in}
\caption{(a) Zigzag map. (b) The profile of entropy-rate for 1,000 Monte Carlo samples from the zigzag map.}
\vspace{-0.15in}
\label{fig:monte-carlo-profile}
\end{figure}

Our results suggest that the entropy-rate (which is a measure of truly randomness) can be used  in the system-level as well as circuit-level optimization of chaotic-map raw-RNGs (along with other measures, such as speed, power consumption, etc).
To better illustrate this, we consider an example using
the zigzag map based TRNG~\cite{ALOG12}, which is a modified version of the tent map, as shown in Fig.~\ref{fig:monte-carlo-profile}(a).
In order to investigate the robustness of the zigzag map to the variations,
we obtained 1,000 Monte Carlo sample maps from the implementation in~\cite{ALOG12}. We further obtained the profile of the entropy-rate of the output bit sequence for the Monte Carlo samples, shown in Fig.~\ref{fig:monte-carlo-profile}(b). The mean of the entropy is 0.97 bits per source symbol suggesting that this raw-RNG is relatively robust to process variations.
To confirm this, we also tested the output bit streams for all the Monte-Carlo samples against NIST 800-22 randomness test suite. Our simulations showed that more than 99\% of the maps generated sequences that passed all the randomness tests (consistent with the very high entropy-rate). Please note that NIST 800-22 tests are computationally very expensive and each test takes minutes to run. Further, NIST 800-22 runs over a sequence generated from a given map, thus, a sequence of 20,000 bits must be generated from the Monte Carlo sample map before the test can be run. On the other hand, given the map function, the entropy-rate calculation takes less than a second and can provide similar predictions. 

\vspace{-.05in}
\subsection{Practical Post-processing Units}
\label{sec:postprocess}
Due to imperfections in practice, the output bits of a raw-RNG are almost always correlated, which calls for the post-processing. Von-Neumann is a widely used post-processing algorithm~\cite{von-neumann51}.
In Von-Neumann algorithm, the sequence is divided into the blocks of length two. Then, $01$ and $10$ are mapped to $0$ and $1$, respectively, while $00$ and $11$ are discarded.
Thus, the Von-Neumann post-processing rate is readily derived to be $R_\text{Von-Neumann} = \frac{1}{2}(\mb{P}[01] + \mb{P}[10])$.
If a truly random bit sequence is fed into the Von-Neumann algorithm, the output rate is only $\frac{1}{4}$, and hence, the Von-Neumann post-processing is not optimal-rate.
For the example map of Fig.~\ref{fig:sample_TCAS2_map}, we have $\mb{P}[01] \approx 0.11$ and $\mb{P}[10] \approx 0.11$. Thus, the output sequence of Von-Neumann post-processing is nearly {\em unbiased} while the post-processing rate $R  \approx 0.11$ is much less than the entropy-rate of $H\approx 0.57$.
Finally, the Von-Neumann algorithm does not generate truly random bits unless the outputs of the raw-RNG are {\em independent}.

Next, we also consider two other post-processing units by Addabbo {\em et. al.}~\cite{addabbo06b} and Poli {\em et. al.}~\cite{Callegari04}. Both papers provide a post-processing unit that results in the generation of bits that pass randomness statistical tests when $H\approx 1$. In both designs, the post-processing rate is equal to one, i.e., $R =1$. Therefore, according to Theorem~\ref{thm:UB}, this post-processing in essence {\em cannot} generate a truly random sequence unless the raw-RNG is a TRNG in the first place.
On other other hand, when the raw-RNG is very close to being truly random, the output sequence of the system has been shown to pass the randomness test suites~\cite{addabbo06b,Callegari04}.
In fact, we applied Addabbo's post-processing to the outputs of the zigzag raw-RNG presented in Section~\ref{subsec:zigzag}, and all of the generated sequences passed NIST 800-22 randomness tests after the post-processing.

We remark that since the condition in Theorem~\ref{thm:LB} is necessary but not sufficient, there are post-processing units that satisfy the rate reduction requirement and still leave patterns in the output stream that are easy to spot using standard randomness tests (e.g., NIST 800-22). On the other hand, although \cite{addabbo06b,Callegari04} do not satisfy the rate reduction condition, they produce output streams that pass randomness test suites.


\vspace{-.07in}
\section{Conclusion}
\vspace{-.03in}
\label{sec:conclusion}
In this paper, we derived the approximate entropy-rate of the hidden Markov process underlying chaotic-map TRNGs.
We concluded that {\em unbiasedness of the generated bits} and the {\em uniformity of steady-state distribution of the chaotic map} are not good measures for the performance evaluation of the TRNG.
Instead,
we proved that the {\em entropy-rate} is the fundamental performance limit of the the bit-generation process from any chaotic-map TRNG and can replace costly NIST~800-22 tests whenever the map function is known.


%


\vspace{-.1in}
\section*{Acknowledgment}
The authors are thankful to the anonymous reviewers whose comments significantly improved the quality of this paper.

\ifCLASSOPTIONcaptionsoff
  \newpage
\fi



%
\vspace{-.09in}
\bibliographystyle{IEEEtran}
\bibliography{TRNG_metric}

\end{document}